\documentclass[a4paper,12pt]{article}
\usepackage{epsfig}
\usepackage[dvips,usenames]{color}
\usepackage{graphicx}

\newlength{\dinwidth}
\newlength{\dinmargin}
\begin{document}
\title{Single production of the vector-like top quark in the littlest Higgs model at TeV energy $e^{-}\gamma$
colliders}
\bigskip
\author{LIU Yao-Bei$^{a}$, WANG Xue-Lei$^{b}$, CAO Yong-Hua$^{a}$\\
{\small a: Henan Institute of Science and Technology, Xinxiang
453003, P.R.China}
\thanks{E-mail:hnxxlyb2000@sina.com}\\
 {\small b: College of Physics and Information
Engineering,}\\
\small{Henan Normal University, Xinxiang  453007, P.R.China}\\
 }
\maketitle
%\date{today}
\begin{abstract}
\indent The existence of a new colored vector-like heavy fermion $T$
is a crucial prediction in little Higgs models, which play a key
role in breaking the electroweak symmetry. The littlest Higgs model
is the most economical one among various little Higgs models. In the
context of the littlest Higgs model, we study single production of
the new heavy vector-like quark and discuss the possibility of
detecting this new particle in the future $LC$ experiment. We find
that the production cross section is in the range of
$1.7\times10^{-3}-30fb$ at TeV energy electron-photon collider with
$\sqrt{s}=3TeV$ and a yearly integrated luminosity of
$\pounds=500fb^{-1}$.
\end{abstract}
PACS number(s): 12.60.Cn, 13.66.Hk, 14.65.Ha
\newpage
\indent The standard model(SM) provides an excellent effective field
theory description of almost all particle physics experiments. But
in the SM the Higgs boson mass suffers from an instability under
radiative corrections. The naturalness argument suggests that the
cutoff scale of the SM is not much above the electroweak scale: New
physics will appear around TeV energies.
 Recently, a new model, known as little Higgs model has drawn a lot of
interest and it offers a very promising solution to the hierarchy
problem in which the Higgs boson is naturally light as a result of
nonlinearly realized symmetry
\cite{little-1,little-2,little-3,littlest}. The key feature of this
model is that the Higgs boson is a pseudo-Goldstone boson of an
approximate global symmetry which is spontaneously broken by a
vacuum expectation value(VEV) at a scale of a few TeV and thus is
naturally light. The most economical little Higgs model is the
so-called
 littlest Higgs model, which is based on a $SU(5)/SO(5)$
 nonlinear sigma model \cite{littlest}. It consists of a $SU(5)$ global
 symmetry, which is spontaneously broken down to $SO(5)$ by a vacuum
 condensate $f$. In this model, a set of new heavy gauge bosons$(B_{H},Z_{H},W_{H})$ and
 a new heavy-vector-like quark(T) are introduced which just cancel
 the quadratic divergence induced by the SM gauge boson loops and the
 top quark loop, respectively. Furthermore, these new particles might produce characteristic signatures
 at the present and future collider experiments \cite{signatures-1,signatures-2,signatures-3}. In little Higgs model, the new heavy vector-like top
 quark $T$ plays a key role in breaking the electroweak symmetry. Thus, studying the possible signatures
 of the new particle $T$ at future high energy colliders would provide significant information
 for $EWSB$ and test the little Higgs mechanism.\\
 \indent It is widely believed that the hadron colliders, such as
 Tevatron and future $LHC$, can directly probe possible new physics
 beyond the standard model(SM) up to a few $TeV$, while the TeV
 energy linear $e^{+}e^{-}$ collider(LC) is also required to
 complement the probe of the new particles with detailed
 measurement \cite{measurement}. An unique feature of the LC is that it can be
 transformed to $\gamma\gamma$ or
 $e\gamma$ colliders with the photon beams generated by laser-scattering
 method. Their effective luminosity and energy are expected to be comparable
 to those of the $LC$. In some scenarios, they are the best
 instrument for the discovery of signatures of new physics. The $e^{-}\gamma$ collisions can produce particles which are
kinematically not accessible in the $e^{+}e^{-}$ collisions at the
same collider \cite{ey}. For example, Ref.\cite{yue} has recently
discussed new single gauge boson production in the littlest Higgs
model at this type of collider.\\
\indent  To avoid the fine tuning problems and produce a suitable
Higgs mass, the mass of the new vector-like top quark $T$ should be
about 2 TeV \cite{littlest}. In this case, the new particle $T$ can
be produced at $LHC$ via two mechanism: $QCD$ pair production via
the processes $gg\rightarrow T\bar{T}$ and $q\bar{q}\rightarrow
T\bar{T}$; single production via $W$ exchange process $qb\rightarrow
q'T$. Due to the large mass of $T$, the later process dominates over
the $QCD$ pair production processes. It has been shown that the new
heavy top quark $T$ mass $M_{T}$ can be explored up to about 2.5TeV
via the $W$ exchange process \cite{signatures-1,W-change}.
Furthermore, Ref.\cite{single-T} has recently discussed single
production of the new heavy vector-like top quark $T$ via the
process $ep\rightarrow eb\rightarrow \nu_{e}T$ at the future
linac-ring type $ep$ collider. In this letter, we will study single
production of the new vector-like top quark $T$ predicted by the LH
model via the process $e^{-}\gamma\rightarrow \nu_{e}b\bar{T}$ and
see whether it can be detected in the future $LC$ experiment with
the c.m. energy $\sqrt{s}=3TeV$ and the integral luminosity
$\pounds=500fb^{-1}$. \\
 \indent The littlest model is
based on the $SU(5)/SO(5)$ nonlinear sigma model. At the scale
$\Lambda_{s}\sim 4\pi$$f$, the global $SU(5)$ symmetry is broken
into its subgroup $SO(5)$ via a vacuum condensate $f$, resulting in
14 Goldstone bosons. The effective field theory of these Goldstone
bosons is parameterized by a non-linear $\sigma$ model with gauged
symmetry $[SU(2)\times U(1)]^{2}$, spontaneously broken down to its
diagonal subgroup $SU(2)\times U(1)$, identified as the SM
electroweak gauge group. Four of these Goldstone bosons are eaten by
the broken gauge generators, leaving 10 states that transform under
the SM gauge group as a doublet H and a triplet $\Phi$. This
breaking scenario also gives rise to four massive gauge bosons
$B_{H}$,$Z_{H}$ and $W^{\pm}_{H}$, A new vector-like top quark $T$
is also needed to cancel the divergence from the top quark loop. All
of these new particles playing together can successfully cancel off
the quadratic
divergence of the Higgs boson mass.\\
\indent In the LH model, the couplings of the heavy vector-like top
quark $T$ to ordinary particles, which are related to our
calculation, can be written as \cite{signatures-1}:
\begin{eqnarray}
g_{V}^{W_{L}e\nu}&=&-g_{A}^{W_{L}e\nu}=\frac{ie}{2\sqrt{2}s_{W}}[1-\frac{v^{2}}
{2f^{2}}c^{2}(c^{2}-s^{2})],\\
g_{V}^{W_{H}e\nu}&=&-g_{A}^{W_{H}e\nu}=-\frac{ie}{2\sqrt{2}s_{W}}\frac{c}{s},\\
g_{V}^{W_{L}Tb}&=&-g_{A}^{W_{L}Tb}=\frac{ie}{2\sqrt{2}s_{W}}\frac{v}
{f}x_{L},\\
g_{V}^{W_{H}Tb}&=&-g_{A}^{W_{H}Tb}=-\frac{ie}{2\sqrt{2}s_{W}}\frac{v}
{f}\frac{c}{s}x_{L}.
\end{eqnarray}
where $f$ is the scalar parameter, $v=246GeV$ is the electroweak
scale, $s_{W}$ represents the sine of the weak mixing angle, and c
is the mixing parameter between $SU(2)_{1}$ and $SU(2)_{2}$ gauge
bosons with $s=\sqrt{1-c^{2}}$. $x_{L}$ is the mixing parameter
between the $SM$ top quark $t$ and the vector-like top quark $T$,
which is defined as
$x_{L}=\lambda_{1}^{2}/(\lambda_{1}^{2}+\lambda_{2}^{2})$,
$\lambda_{1}$ and $\lambda_{2}$ are the Yukawa couplings parameters.
We write the gauge boson-fermion couplings in the form of
  $i\gamma^{\mu}(g_{V}+g_{A}\gamma^{5})$. The mass of the heavy
vector-like top quark $T$ can be approximately written as:
\begin{eqnarray}
M_{T}&=&\frac{m_{t}f}{v}\sqrt{\frac{1}{x_{L}(1-x_{L})}}[1-\frac{v^{2}}{2f^{2}}x_{L}(1+x_{L})].
\end{eqnarray}
 \indent  The number of up-type quarks is four in the LH model and thus the matrix
 relating the quark mass eigenstates with the weak eigenstates becomes a $4\times3$ matrix. Compared to the
 $CKM$
 matrix in the SM, the extended $CKM$ matrix has the fourth row elements $V_{Td}$, $V_{Ts}$ and $V_{Tb}$. Thus,
  it is possible that there are the decay channels $T\rightarrow ql\bar{\nu}$, which q is the down-type quark. However, their
  branching ratio are very small \cite{small}. Thus, the dominant decay modes of the heavy vector-like top quark $T$
 are $tH$, $tZ$,and $bW$ with partial widths in the ratio 1:1:2
 \cite{signatures-1,ratio}. At the order of $v^{2}/f^{2}$, the total width of the
 new vector-like top quark $T$ can be written as:
\begin{eqnarray}
\Gamma_{T}=\frac{M_{T}}{8\pi}(\frac{m_{t}}{v})^{2}\frac{x_{L}}{1-x_{L}}.
\end{eqnarray}
\begin{figure}[t]
\begin{center}
\epsfig{file=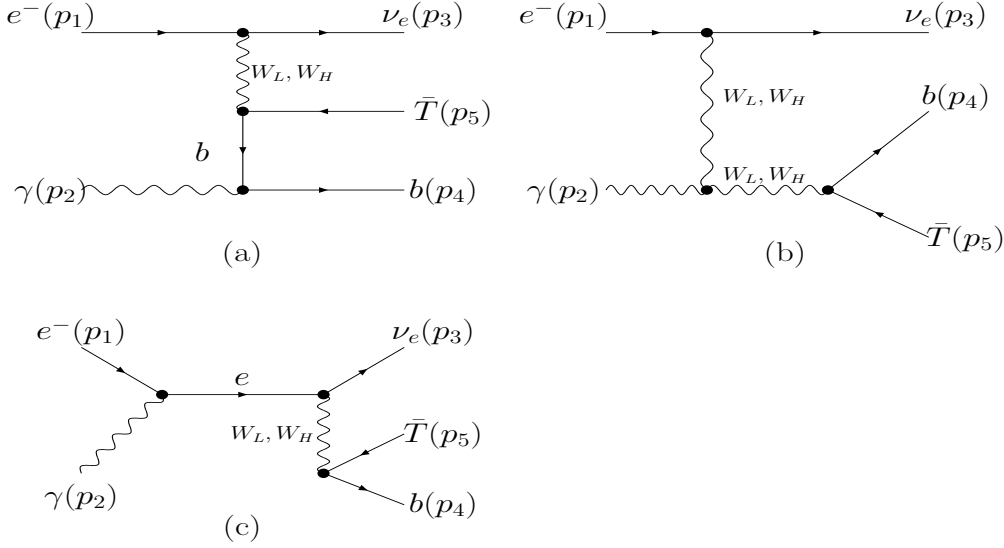,width=450pt,height=500pt} \vspace{-9cm}
\caption{\small Feynman diagrams of the process
$e^{-}\gamma\rightarrow \nu_{e}b\bar{T}$ in the littlest Higgs
model.} \label{fig1}
\end{center}
\end{figure}
\indent  In this case, we can see that the new heavy vector-like top
quark $T$ can be single produced via the process
$e^{-}\gamma\rightarrow \nu_{e}b\bar{T}$ via $e^{-}\gamma$
collisions in the future $LC$ experiment with $\sqrt{s}=3TeV$ and
$\pounds=500fb^{-1}$.
 The relevant tree-level Feynman diagrams of the process are shown in Fig.1.
\indent In order to write a compact expression for the amplitudes,
it is
 necessary to define the triple-boson couplings coefficient as:
\begin{equation}
 \Gamma^{\alpha\beta\gamma}(p_{1},p_{2},p_{3})=g^{\alpha\beta}(p_{1}-p_{2})^{\gamma}
 +g^{\beta\gamma}(p_{2}-p_{3})^{\alpha}+g^{\gamma\alpha}(p_{3}-p_{1})^{\beta},
 \end{equation}
 with all motenta out-going.
The invariant production amplitudes of the process
 can be written as:
\begin{equation}
 M=M_{a}+M_{b}+M_{c}
 \end{equation}
 with
 \begin{eqnarray*}
 M_{a}&=&-\frac{ie^{3}}{8s_{W}^{2}}\frac{v}{f}x_{L}\bar{u}(p_{3})\gamma_{\mu}(1-\gamma_{5})u(p_{1})\{
 G(p_{3}-p_{1},M_{W})+\frac{c^{2}}{s^{2}}G(p_{3}-p_{1},M_{W_{H}})\}\nonumber\\
& &\times g^{\mu\nu}\bar{u}(p_{4})\gamma_{\rho}G(p_{4}-p_{2},m_{b})
 \gamma_{\nu}(1-\gamma_{5})v(p_{5})\varepsilon^{\rho}(p_{2}),\\
 M_{b}&=&\frac{ie^{3}}{8s_{W}^{2}}\frac{v}{f}x_{L}\bar{u}(p_{3})\gamma_{\mu}(1-\gamma_{5})u(p_{1})
 \Gamma^{\mu\nu\rho}(p_{3}-p_{1},-p_{2},p_{4}+p_{5})\{G(p_{3}-p_{1},M_{W})\nonumber\\
 & &\times G(p_{4}+p_{5},M_{W})+\frac{c^{2}}{s^{2}}G(p_{3}-p_{1},M_{W_{H}})G(p_{4}+p_{5},M_{W_{H}})\}
\bar{u}(p_{4})\gamma_{\nu}(1-\gamma_{5})v(p_{5})\varepsilon^{\rho}(p_{2}),\\
M_{c}&=&-\frac{ie^{3}}{8s_{W}^{2}}\frac{v}{f}x_{L}\bar{u}(p_{4})\gamma_{\mu}(1-\gamma_{5})v(p_{5})\{
 G(p_{4}+p_{5},M_{W})+\frac{c^{2}}{s^{2}}G(p_{4}+p_{5},M_{W_{H}})\}\nonumber\\
 & &\times g^{\mu\nu}\bar{u}(p_{3})\gamma_{\nu}(1-\gamma_{5})G(p_{1}+p_{2},0)
 \gamma_{\rho}u(p_{1})\varepsilon^{\rho}(p_{2}).
  \end{eqnarray*}
 where $G(p,m)=1/(p^{2}-m^{2})$ denotes the propagator of the
 particle.\\
\indent The hard photon beam of the $e\gamma$ collider can be
obtained from laser backscattering at the $e^{+}e^{-}$ linear
collider. Let $\hat{s}$ and $s$ be the center-of-mass energies of
the $e\gamma$ and $e^{+}e^{-}$ systems, respectively. After
calculating the cross section $\sigma(\hat{s})$ for the subprocess
$e^{-}\gamma\rightarrow \nu_{e}b\bar{T}$, the total cross section at
the $e^{+}e^{-}$ linear collider can be obtained by folding
$\sigma(\hat{s})$ with the photon distribution function that is
given in Ref\cite{function}:
\begin{equation}
\sigma(tot)=\int^{x_{max}}_{(M_{T}+M_{b})^{2}/s}dx\sigma(\hat{s})f_{\gamma}(x),
 \end{equation}
where
\begin{equation}
f_{\gamma}(x)=\frac{1}{D(\xi)}[1-x+\frac{1}{1-x}-\frac{4x}{\xi(1-x)}+\frac{4x^{2}}{\xi^{2}(1-x)^{2}}],
\end{equation}
with
\begin{equation}
D(\xi)=(1-\frac{4}{\xi}-\frac{8}{\xi^{2}})\ln(1+\xi)+\frac{1}{2}+\frac{8}{\xi}-\frac{1}{2(1+\xi)^{2}},
\end{equation}
In the above equation, $\xi=4E_{e}\omega_{0}/m_{e}^{2}$ in which
$m_{e}$ and $E_{e}$ stand, respectively, for the incident electron
mass and energy, $\omega_{0}$ stands for the laser photon energy,
and $x=\omega/E_{e}$ stands for the fraction of energy of the
incident electron carried by the backscattered photon. $f_{\gamma}$
vanishes for $x>x_{max}=\omega_{max}/E_{e}=\xi/(1+\xi)$. In order to
avoid the creation of $e^{+}e^{-}$ pairs by the interaction of the
incident and backscattered photons, we require
$\omega_{0}x_{max}\leq m_{e}^{2}/E_{e}$, which implies that $\xi\leq
2+2\sqrt{2}\simeq4.8$. For the choice of $\xi=4.8$, we obtain
\begin{equation}
x_{max}\approx0.83,~~~~~~~~~~~~~~~~~~~D(\xi_{max})\approx1.8.
 \end{equation}
 For simplicity, we have ignored the possible polarization for the
 electron and photon beams.\\
\indent With above production amplitudes, we can obtain the
production cross section directly. In the calculation of the cross
section, instead of calculating the square of the amplitudes
analytically, we calculate the amplitudes numerically by using the
method of the references \cite{HZ} which can greatly simplify our
calculation. \\
\indent From the above discussion, one can see that
single $T$ production in the channel $e^{-}\gamma\rightarrow
\nu_{e}b\bar{T}$ at TeV energy $LC$ colliders comes from two
processes: the $SM$ gauge boson $W$ exchange and the new gauge boson
$W_{H}$ exchange. The contributions of the former process mainly
depend on the free parameters $M_{T}$ and $x_{L}$, while those of
the latter process mainly dependent on the free parameters $M_{T}$,
$x_{L}$, c and $M_{W_{H}}$. Taking into account the precision
electroweak constraint on the parameter space of the $LH$ model, the
free parameters $x_{L}$, c and $M_{W_{H}}$ are allowed in the ranges
of $0<x_{L}<1$, $0\leq c\leq 0.5$, and $1TeV\leq M_{W_{H}}\leq3TeV$
\cite{parameters}. Observably, the cross section of single $T$
production mainly comes from $W$ exchange and is not sensitive to
the free parameters c and the $W_{H}$ mass $M_{W_{H}}$. So, we
will take c=0.3 and $M_{W_{H}}=2TeV$ in our numerical calculation.\\
%%%%%%%%%%%%%%%%%%%%%%%%%%%%%%%%%%%%%%%%%%%%%%%%%%%%%%%%%%%%%%%%%
\begin{figure}[h]
\begin{center}
\scalebox{0.9}{\epsfig{file=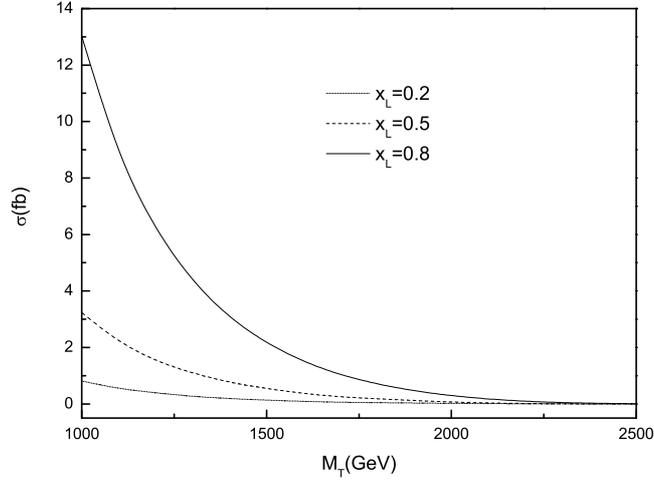}}\\
\end{center}
\caption{\small The cross section $\sigma(s)$ of single T production
as a function of $M_{T}$ for c=0.3 and three values of the parameter
$x_{L}$.}
\end{figure}
\indent In Fig.2, we plot the single production section $\sigma(s)$
of the heavy vector-like top quark $T$ as a function of $T$ quark
mass $M_{T}$ for three values of the mixing parameter $x_{L}$. One
can see from Fig.2 that the cross section $\sigma(s)$ fall sharply
with $M_{T}$ increasing and increases as the mixing parameter
$x_{L}$ increases. For $x_{L}$=0.5, and $1TeV\leq M_{T}\leq2.5TeV$,
the value of the cross section $\sigma(s)$ is in the range of
$3.24fb\sim 1.7\times10^{-3}fb$. If we assume that the yearly
integrated luminosity $\pounds=500fb^{-1}$, then there could be as
much as 1500 events produced each year.   \\
\begin{figure}[h]
\begin{center}
\scalebox{0.9}{\epsfig{file=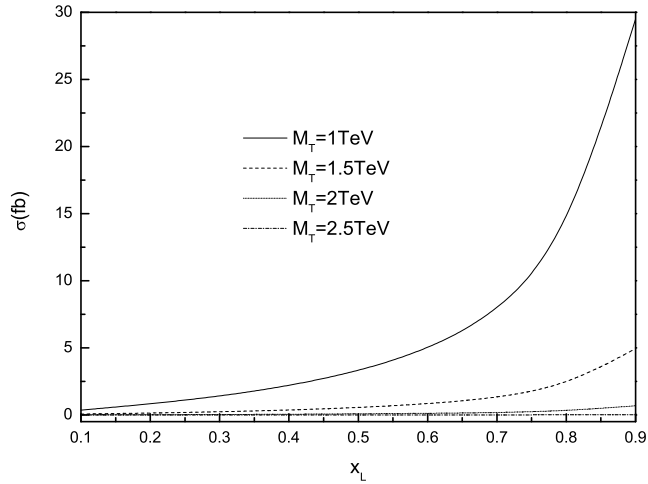}}
\end{center}
\caption{\small The cross section $\sigma(s)$ of single T production
as a function of $x_{L}$ for different values of $M_{T}$.}
\end{figure}
\indent To see the influence of the mixing parameter $x_{L}$ on the
cross section, in Fig.3, we plot $\sigma(s)$ as a function of the
mixing parameter $x_{L}$ for $M_{T}=1 TeV, 1.5TeV,2TeV,2.5TeV$,
respectively. From Fig.3, one can see that the cross section
increases as the mixing parameter $x_{L}$ increasing for fixed $T$
quark mass $M_{T}$. This is because the value of the production
section is proportional to the mixing parameter $x_{L}$ which comes
from the flavor change couplings $g^{W\bar{T}b}$. On the other hand,
as long as the heavy vector-like top quark $T$ mass $M_{T}$ is
smaller than 2TeV, the cross section can even reach tens $fb$. If we
take integral luminosity $\pounds=500fb^{-1}$, there are about
$10^{3}-10^{4}$ $\nu_{e}b\bar{T}$ events to be produced. There will
be a promising number of fully reconstructible events to detect
single $T$ production via $e^{-}\gamma$
collisions in future $LC$ experiment with $\sqrt{s}=3TeV$ and $\pounds=500fb^{-1}$. \\
 \indent Little Higgs theory has generated much interest as one kind
 of models of $EWSB$, which can be regarded as one of the important
 candidates of new physics beyond the $SM$. For all of the little
 Higgs models, at least one vector-like top quark $T$ is needed to
 cancel the numerically most large quadratic divergence coming from
 top Yukawa couplings. At the leading order, the heavy vector-like top
 quark $T$ predicted by the $LH$ model mainly decays to the $tZ$,
 $tH$
 and $bW$ modes, which can provide characteristic signatures for the
 discovery of the heavy vector-like quark $T$ in the future high
 energy collider experiments. It has been shown that the signal of
 the new vector-like quark $T$ might be detected via all of the
 three decay modes in the future $LHC$ experiment and the linac-ring type
 $ep$ collider. In this paper, we study single
 heavy vector-like top quark $T$ production via the process $e^{-}\gamma\rightarrow \nu_{e}b\bar{T}$
 in the future high energy $LC$ experiment. With the favorable parameter
 spaces, the sufficient events can be produced to detected the
 single of the vector-like top quark $T$ at the $TeV$ energy
 $e^{-}\gamma$ colliders.

\end{document}